\documentclass[12pt]{article}
 \usepackage{graphicx,psfrag,epsfig}

\makeatletter
\def\ifempty#1{\@ifempty #1\@emptymarkA\@emptymarkB}%
\def\@ifempty#1#2\@emptymarkB{\ifx #1\emptymarkA}%
\def\@emptymarkA{\@emptymarkA}%

\renewcommand{\part}[1]{%
    \stepcounter{part}%
    \begin{center}%
    {\large\bfseries\thepart.\ #1}
    \end{center}}%

\renewcommand{\@seccntformat}[1]{%
    {\csname the#1\endcsname}. \ 
    }

\renewcommand{\section}{%
    \@startsection{section}{1}{\z@}%
    {-3.5ex plus -1ex minus -.2ex}%
    {2.3ex plus.2ex}%
    {\centering\large\bfseries}}

\renewcommand{\subsection}{\@startsection{subsection}{2}{0pt}%
    {-3.25ex plus -1ex minus -.2ex}%
    {1.5ex plus .2ex}%
    {\centering\normalsize\bfseries}}

\renewcommand{\subsubsection}{\@startsection{subsubsection}{2}{0pt}%
    {-3.25ex plus -1ex minus -.2ex}%
    {1.5ex plus .2ex}%
    {\centering\normalsize\itshape}}

\renewenvironment{thebibliography}[1]
     {\part{REFERENCES CITED}%
      \list{\@biblabel{\@arabic\c@enumiv}}%
           {\settowidth\labelwidth{\@biblabel{#1}}%
            \leftmargin\labelwidth
            \advance\leftmargin\labelsep
            \@openbib@code
            \usecounter{enumiv}%
            \let\p@enumiv\@empty
            \renewcommand\theenumiv{\@arabic\c@enumiv}}%
      \sloppy\clubpenalty4000\widowpenalty4000%
      \sfcode`\.\@m}
     {\def\@noitemerr
       {\@latex@warning{Empty `thebibliography' environment}}%
      \endlist}

\@addtoreset{page}{chapter}
\def\@stpelt#1{\global\csname c@#1\endcsname%
    \expandafter\ifx \csname#1\endcsname \page%
        \@ne%
    \else%
        \z@ \fi}

\makeatother

\begin{document}
\begin{center}
{\Large \bf Weak Interactions: From Current-Current to Standard Model and Beyond\footnote{Versions of this article will appear as a chapter in the book ``Hundred years of sub-atomic physics" edited by Ernest Henley, to be published by World Scientific Publishing and in International Journal of Theoretical Physics, WSPC (2012) .}}\end{center}

\begin{center}
 {Rabindra N. Mohapatra}\\
{\it Maryland
Center for Fundamental Physics and Department of Physics,
University of Maryland, College Park, Maryland 20742, USA}
\end{center}
\date{\today}
\begin{abstract}
This paper provides a brief overview  for non-specialists of  some of the highlights in the development of the theory of weak interactions  during the past century.
\end{abstract}


\section{Early history}
In 1896, French physicist Henri Bacquerel made the discovery of a new kind of spontaneous radiation from Uranium salt that 
 formed the foundation for a whole new field of research in subatomic physics. It held not only  far reaching implications for our understanding  of the
 fundamental forces of nature, but also resolved  many puzzles in astrophysics and cosmology. The Bacquerel rays (also known as beta rays) were nothing other than
 electrons emitted in the spontaneous disintegration of Uranium nuclei, a phenomenon which was subsequently found  to be replicated in many other nuclear isotopes
 establishing thereby a whole new class of phenomena in nature. The slowness of
 beta emission, a characteristic feature of all nuclear beta ray emission led to the realization that the strength of the force responsible for this process
  was much smaller than that of the electric force responsible for binding electrons in atoms or the nuclear force (also known as the strong force) that binds nucleons to form nuclei;  hence the name, weak interactions (or weak force) to describe this new class of phenomena. The weak force together with the three other forces
 of nature, the strong, electromagnetic and gravitational,  make up the four forces known today. The present article is a pedagogical overview
 of the developments that culminated in a successful theoretical  understanding of the weak force, the questions raised by related developments, and outlook for
 what may lie ahead in this field of study in decades to come.
 
 A major step that led to the understanding of the nature of force responsible for beta rays came from the observation of Chadwick about 20 years later that
 there was a spread in the energy of the emitted electrons,  apparently contradicting the conservation of energy if only  electrons were emitted in the process. The reason is that if only an electron had been emitted in beta decay, 
 energy conservation would have implied that it  be mono-energetic, with energy equal to the difference between the masses of the initial and final nuclei. This puzzle of a 
 continuous energy spectrum was solved by Wolfgang Pauli in 1930 when he wrote  a letter to his colleagues gathered at a conference that he could not attend addressing them as ``Dear radioactive Ladies and Gentlemen" and suggested 
 that  in beta emission from nuclei, the electrons were always accompanied by a tiny electrically-neutral, spin half, massive particle called the neutrino. (Pauli in
 his letter, written prior to the discovery of the neutron called it ``neutron" and Fermi changed it to ``neutrino''). The neutrino was discovered some quarter century later
 by Reines and Cowan and the rest of the history of weak interactions could arguably be thought of as our attempts to understand the neutrino, although admittedly, there was more.
 
 As neutrinos became the  focus of discussion in weak interactions, new developments followed. A major breakthrough  came not only in our understanding of the neutrino but also the nature of the weak force  in 1956 when Lee and Yang proposed that weak forces do not obey symmetry (``parity'')  under mirror reflection~\cite{leeyang}. This came about as they were attempting to
 understand the so-called $\tau-\theta$ puzzle . This suggestion was revolutionary since until that time, all
 forces of nature were  believed to respect mirror reflection symmetry. The  $\tau-\theta$ puzzle involved  two particles (called $\tau$ and $\theta$) with identical properties e.g. mass $M$ about 500 MeV and charge and spin zero,  were found to decay to different final states, one containing two  pions and the other three  pions. Since the $\pi$-mesons were known to have odd parity and their orbital angular momenta were consistent with $\ell =0$ (reasonable since $M$ is quite small), the two pion state would have even parity whereas the three pion state would have odd parity. The identical properties of the two decaying particles  suggested that they are one and the same whereas if parity conservation is a good symmetry, they could not correspond to the same particle. This was the puzzle. 
 The Lee-Yang resolution was that there is only one particle but  that parity  invariance does not hold for weak forces  which causes the decay of the particle thought to be $\tau$ and $\theta$. Indeed, it was soon realized that the particle is none other than the neutral K-meson. The Lee-Yang conjecture
 was confirmed the following year in an experiment by Ambler, Hayward, Hudson and Wu~\cite{wu} who observed asymmetric emission of electrons from polarized Co$^{60}$ nuclei
 as would be the case if weak interaction violated parity (as opposed to having symmetric emission if parity were a good symmetry of weak forces). This revolutionary suggestion implied that the weak interaction  was ``the odd man out'' among the forces of nature.
 
 In parallel with the above developments, our understanding of fundamental building blocks of nature was also slowly expanding. More and more ``elementary" particles were being discovered: in addition to protons, neutrons and electrons which were already known by the 1930s, new particles e.g. $\pi^{\pm}$, $\pi^0$,
 $K^{\pm}$, $K^0$, $\mu$ etc were discovered from studies of cosmic rays as well as at accelerators. These were classified into hadrons and leptons depending on what kind of interactions they participated in: hadrons if they had strong as well as weak and electromagnetic interactions  (e.g. $p, n, \pi, K$) and leptons if they did not have  strong interactions but participated in the rest of the interactions. All particle are known to have gravitational interactions.
 
To categorize the behavior of the particles in production and decay, they were assigned quantum numbers: for instance, stability of the hydrogen atom could be understood if the proton and electron were given a baryon number (B=1) and a lepton number (L=1)  respectively. Similarly since two kinds of K-mesons (a $K$ and  $\bar{K}$) were produced in  strong interaction processes , they were assigned a new quantum number called strangeness (S). To be consistent with observations, strong and electromagnetic forces would be  expected to conserve all these quantum numbers whereas other weak interactions possibly would not. Such quantum numbers would  signify the presence of some higher symmetry- for instance, strangeness as part of an $SU(3)$ symmetry for hadrons that we mention below.
 
 \section{From Four-Fermion Interaction to V-A theory}
 Soon after Pauli's suggestion of a neutrino, Fermi proposed the four-fermion theory of weak interactions. His proposal was deeply rooted in the framework of quantum field theory 
 where particle production is more natural (unlike quantum mechanics) and described by the annihilation of a neutron inside a nucleus  with simultaneous appearance of a proton, an electron and an anti-neutrino due through the four-fermion interaction. This process happens with a certain strength, the Fermi constant, given by $G_F$, which is a small dimensioned number that accomodates the long life times of the weak-decay processes. The Hamiltonian for the four-fermion interaction can be written as
 \begin{eqnarray}
 {\cal H}_{wk}~=~\frac{G_F}{\sqrt{2}}\bar{\psi}_p\psi_n\bar{\psi}_e\psi_\nu + h.c.
 \end{eqnarray}
 where each field operator is responsible for the creation (of $\bar{\psi}$) or destruction (of $\psi$) of the particle that appears as the subscript.
 This generic form  accounted for processes such as the usual nuclear beta decay where $n\to p+e^-+\bar{\nu}_e$ as well as for muon decay $\mu^-\to e^-+\nu_\mu+\bar{\nu}_e$  (with corresponding particle labels changed) with overall strength of coupling $G_F\approx 10^{-5}$ GeV$^{-2}$. The discovery that both processes had nearly the same strength provided the insight of their being a  universality for weak interactions. This was the key observation that led to the birth of modern gauge theories as will be described below. 
 
 An important property of the current-current form of weak interaction for leptons is that it provides a way for assigning a lepton number to $e, \nu_e, \mu, \nu_\mu, \tau, \nu_{\tau}$  that is conserved not only for the electromagnetic forces but also for the weak  current-current Hamiltonian. We will see shortly that this quantum number has a central role in the  developments of the field of weak interactions. 
 
 While the generic four-fermion theory captured the overall feature of weak processes, further progress demanded greater understanding  of the details of the nature of the interaction. The point is that fundamentals of quantum mechanics suggests that  fermion bilinear terms can have different Lorentz structure: scalar, pseudoscalar, vector, axial vector and tensor, each with characteristic predictions for experiments . The question is: which of these play a role in weak interaction  ? 
 
  The years following the discovery of parity violation were exceptionally interesting  involving a symbiotic interplay between experiment and theory that eventually  settled this mystery, when Feynman, Gell-Mann, Sudarshan and Marshak~\cite{marshaksudarshan}  proposed in 1957 that, of all the different Lorentz possibilities, it was the V-A combination for each set of fermions that forms the four-fermion interaction. This was a particularly bold suggestion since at that time experiments seemed to contradict this possibility. The V-A option also incorporated the possibility that  parity is violated in weak interactions.  Stunning experimental support for this theory came from the discovery by Goldhaber, Grodzins and Sunyar~\cite{ggs} that neutrinos emitted in weak processes were left handed as  predicted by the V-A theory. Other experiments soon followed  and the V-A theory became accepted as the fundamental basis for all weak interactions..
 
 The next phase of the development  in the late 1950s and early 1960s focussed on gaining an understanding of all weak interactions in terms of the four-fermion V-A picture, which required  a systematic characterization of all weak processes. There seemed to be three classes of interactions: (i) purely leptonic processes, that involved only leptons which do not have strong interactions; (ii) semi-leptonic processes that involved both leptons and  hadrons (e.g. $n\to pe^- \bar{\nu}$), ($\pi^+ ~or ~K^+\to e^+\nu$) and finally (iii) non-leptonic interactions that did not involve any leptons e.g. $(K^0\to \pi^+\pi^-, \pi^0\pi^0, \Lambda\to p \pi^-)$ etc. Characteristics of the purely leptonic decays were the easiest to calculate using perturbative quantum field theory as they were free from the complications of strong interaction and the weak force had far too weak a coupling for higher order processes to have significant effects (at least at low energies). The other class of processes needed additional insight before reliable calculations could be formulated to check the validity of the  theory. 
 
 The key question was how to define the fermion bilinears when they involved hadrons and how to compute the weak processes that involved strongly interacting particles such as protons, neutrons and pions. Analogy with electromagnetic interactions as embodied in the successful quantum electrodynamics (QED) became the guiding principle. The work of Schwinger, Feynman and Tomonaga had by then  led to a theory of all electromagnetic processes, the key ingredient of which was an interaction that provided  a vector bilinear term for electrically charged fields such as the electron, muon, proton with the photon field.  The weak interactions on the other hand involved the product of two V-A  currents. The bilinear and vector nature of both the electromagnetic and weak currents was unmistakably suggestive  of a fundamental unity between weak and electromagnetic interactions. This was exploited by Gell-Mann and others to propose the conserved vector current hypothesis (CVC) for the vector part of the hadronic weak currents as well as the hypothesis of partial conservation of the axial part of the hadronic weak currents (PCAC). According to these postulates, the hadronic current (both V and the A parts) was a close ``cousins'' of the familiar electric current and both were expected to have similar properties. For example, the electric current is a conserved current so that when its time component is integrated over all space, it yields the electric charge and this property makes it easier to study the electromagnetic scattering of hadrons. The question then arose whether the weak current   behaved and there is some analog of the electromagnetic charge for an  integration of the time component of the weak current over all space ?
  
 \section{Symmetries become the dominant theme}
 To understand how the above question was answered, it will help to discuss the dominant approach to particle interactions in the late fifties and  sixties. The near equality of the mass of the proton and the neutron had prompted Heisenberg to suggest the idea that there must be a symmetry of strong interactions under which the proton transforms to the neutron, since the belief was that most of the mass of the protons and neutrons came from the strong force. This symmetry was called strong isospin and  Noether's theorem implied therefore the existence of  conserved currents  whose time component  integrated over all space generated the isospin  ($I$) symmetry.   Isospin is like spin, except that it operates in an internal space. The closeness of the  mass values of the proton and neutron was interpreted in terms of their being the "up" and "down"  components of an $I=\frac{1}{2}$ representation.  Gell-Mann suggested that, in analogy, the vector part of the weak current corresponds to a conserved weak isospin current whose time component when integrated over all space also generates a weak isospin symmetry. So the answer to the question that ended the last section came therefore  from  the unlikely domain of strong interactions.  
 
 What about the axial weak current that arises in a V-A description of weak interactions ? Could that be the analog of the generator of a strong interaction symmetry just like the vector current ? Since the parity of the axial current is different from that of the vector current,  a strong axial isospin symmetry would require that there be a parity odd partner of the proton and neutron, for which there is no evidence.  Gell-Mann therefore made the bold suggestion that axial current corresponds to only an approximate symmetry and is therefore  only approximately conserved. This was called the  PCAC hypothesis (or partially conserved axial current). It is certainly interesting that there is no parity doubling of the observed hadronic spectrum i.e. there were no odd parity partners for protons, neutrons etc. 
So, how such a symmetry is realized in nature is a question that needed to be answered and as we will see below understanding this  led to the ground breaking works of Nambu and  Goldstone who proposed that not all symmetries are realized through multiplets of nearly same mass but  could be spontaneously broken by the fact that the ground state of the theory does not respect the symmetry. This insight had profound implications as discussed below.  

 With the introduction of vector isospin as an approximate symmetry and possibly also axial isospin as a symmetry and axial isospin as an approximate symmetry group theoretic discussions were slowly creeping into particle physics. In the early sixties, Gell-Mann and Ne'eman extended Heisenberg's idea to introduce  $SU(3)$ as an approximate symmetry of the hadron spectrum that allowed the symmetry concept to cover not just proton, neutron and pion but to much more "exotic" hadrons of that time the $\Sigma$ , $\Xi$, K mesons etc. According to the ideas of group theory, if these symmetries are exact, their generators are time independent and form a Lie algebra. The algebras have irreducible representations to which particles can be assigned and  properties such as their seemingly unrelated couplings can be related to one  another . In the limit that the symmetries are exact, masses of all the particles in a given irreducible representation of the symmetry would be exact. 
However, the masses of proton and neutron (which are approximately equal up to small electromagnetic corrections) are different from those of $\Lambda$, $\Sigma$  and $\Xi$ which are supposed to be part of the same irreducible representation.  The question now is : how to treat the effects of strong interaction  on calculations based on a broken symmetry. Since the strong coupling is large, it is not wise to rely on perturbative expansion.  Nevertheless group theory is very successful in determining masses as was shown by the success of the Gell-Mann-Okubo mass formula for baryon and meson multiplets. For weak interactions this kind of procedure is not useful and requires an alternative approach.

To get around this difficulty,  Gell-Mann proposed what is known as current algebra which posits that the same group (e.g.  SU(3)), Lie algebra is still satisfied but not just by the generators, which for broken symmetry are not time independent, but by currents as well as the time-dependent generators. Their approximate conservation property can be used to carry out  calculations of  weak processes involving certain hadrons. This opened up the field to many calculational opportunities. The most celebrated  is the Adler-Weisberger calculation of the axial coupling $g_A$ for weak interactions which provides the first reliable way to estimate the lifetimes for neutron beta decay. Many other applications followed.

\subsection{From symmetries to quarks to symmetries}
 
 The current-current form  for the weak interaction  Hamiltonian can now be written as:
 \begin{eqnarray}
 {\cal H}_{wk} ~ =~\frac{G_F}{\sqrt{2}}{\cal J}^\alpha {\cal J}_\alpha + h.c.
 \end{eqnarray}
 where ${\cal J}_\alpha = \bar{\psi}_e\gamma_\alpha(1-\gamma_5)\psi_{\nu_e} + J_\alpha$ , $J_\alpha= V_\alpha - A_\alpha$ being the difference between the vector and axial vector hadronic currents respectively. The strong interactions effects could then be treated in a nonperturbative manner within the framework of the current current theory. 
 
 Another aspect of weak interaction is the way  strange hadrons (i.e. the hadrons such as $K, \Lambda, \Sigma...$ etc) participate in weak processes. Their weak interaction rates are found  to be weaker than for normal hadrons such as neutrons or pions. To accommodate this difference,  Cabibbo proposed that in the weak interaction Hamiltonian must contain a mixing between the non-strange and strange hadrons with the weak current involving hadrons  written as $\bar{p}\gamma^\mu (1-\gamma_5)(n \cos\theta+\Lambda \sin\theta)$ where $\theta$ is  known as the Cabibbo angle. The value of the Cabibbo angle is about $13^0$. The same angle also accommodates the weak decays involving strange mesons such as the kaon. 
 
 In 1964, Gell-Mann and Zweig~\cite{gz} proposed that all hadrons (protons, neutrons, pions etc) were made of more fundamental constituents called quarks. The hadrons were nothing but quarks bound by the strong force. According to the quark model, the nucleon was made of  $u$ and $d$ quarks with $p\equiv uud$ and $n=udd$ etc. As evidence for quark substructure of hadrons grew,  it also impacted  the weak interactions with the charged hadronic weak currents $V_\alpha$ and $A_\alpha$ expressed in terms of  quarks  as  $V_\alpha = \bar{u}\gamma_\alpha d$ and $A_\alpha=  \bar{u}\gamma_\alpha \gamma_5 d$. This form implied that as far as the weak forces go, there is an amazing similarity between the neutrino 
( $\nu$)  and the up quark ( $u$)  and between the electron ( $e$) and the down quark  ($d$). Since $u$ and $d$ correspond to the  up and down components of the strong interaction symmetry group isospin, one could surmise that there is a weak isospin group whose up and down components are $\nu$ and $e$ as well as $u$ and $d$. More precisely, since the weak current is pure left-handed or (V-A) type, one could write left handed helicity parts of these fields as the components of weak isospin doublet. Includinng the strange quark, the weak current can be written in terms of quark bilinears as $\bar{u}\gamma^\mu (1-\gamma_5)(d \cos\theta+s \sin\theta)$ .
 
 By the late fifties, symmetries were considered a fundamental part of our understanding of strong forces and hadron spectrum. The quark model emerged as a way to handle the strong interaction symmetries. Writing the weak currents in terms of quark bilinears provided a clue that there may be hidden symmetries for weak forces too, an intuitive notion that eventually led to the incredible journey from the current-current form for weak interactions to the standard model (SM) of weak forces, every detail of which seems to have been confirmed, first by the discovery of weak neutral currents, the W and Z bosons\cite{WZ} and finally this year, the discovery of the Higgs boson\cite{higgs} (although in the last case how accurately the observed boson represents the  SM Higgs rather than some beyond the standard model field is not clear yet.) 
 
 \subsection{Enter the charm quark} 
 A significant milestone in the history of weak interaction is the discovery of the charm quark. In 1959, Gamba, Marshak and Okubo~\cite{gmo} suggested the idea of quark lepton symmetry in weak interactions. They observed the correspondence between $p\leftrightarrow \nu$, $n\leftrightarrow e$ and $\Lambda\leftrightarrow \mu$. This was before the discovery of the muon neutrino ( $\nu_\mu$),  by Lederman, Steinberger and Schwarz\cite{ls}. In 1962, in order to reconcile the idea of quark-lepton symmetry  with the fourth lepton $\nu_\mu$, Maki, Nakagawa and Sakata~\cite{mns} suggested in a seminal paper  that  neutrinos mix among themselves. True reconciliation between quark lepton symmetry and the four leptons came in the paper by Bjorken and Glashow~\cite{bg}, who proposed that  the Gamba-Marshak-Okubo hypothesis required a fourth quark, the charm quark $c$  to retain the $p\leftrightarrow \nu_e $ and $c\leftrightarrow \nu_\mu$ correspondence. The true significance of the role of the charm quark came a few years later, which is in itself  an interesting history.
 
 As the current-current form became the cornerstone for describing the weak interactions, its field theoretic underpinnings started to concern the particle theorists. QED was  established as a remarkably successful theory of electrons and photons, where one could calculate quantum corrections such as Lamb shift, g-2 of electrons, both in agreement with observations. Such calculations were reliable and their comparision with experiment was possible because QED was a renormalizable theory which had only interactions with mass dimension four or less. The burning question in late sixties was ``is there also a renormalizable theory of weak interactions  ?" The current-current theory based on the four fermion interaction of dimension six was clearly not renormalizable and led to infinite integrals whenever high loop corrections were calculated. Many attempts were made in the mid to late sixties to make sense of a theory that was so successful in describing observations and yet was plagued with infinities ?  One approach to doing this was simply to put a cut-off to make sense of these infinities. It was however pointed out in a series of papers ~\cite{mmris} that when such cut-off procedure was applied to calculating strangeness changing processes such as $K_L\to \mu^+\mu^-$ and $K_L-K_S$ mass differences in the four-fermion theory, observations required that the needed cutoff was low , near few GeVs. One way to interpret such a low cut-off was to assume that the cut-off originated from new physics near a few GeVs.  This prompted Glashow, Iliopoulos and Maiani~\cite{GIM} to suggest that this cut-off can be identified with the mass of the Bjorken-Glashow charm quark, which then led to a clarification of the magnitude of higher order contributions from weak processes. This came to be known as the  GIM mechanism. The charm particle was eventually discovered in 1974 by two experiments, one at Brookhaven by the group of Sam Ting\cite{ting} and the other at SLAC by the group of Burton Richter\cite{richter}, both of whom were awarded the Nobel prize for this discovery.
 
\section{Birth of the standard model}

The end of the 1960s was a chaotic period in particle physics. The three major fundamental forces of concern to particle physicists  were viewed in three different ways. As noted earlier, QED was considered ``in the bag" in the sense that viewed as a theory describing the electromagnetic interaction  of particles such as electrons, muons etc, it was fully understood in terms of a renormalizable field theory whose Lagrangian was determined by invariance under a local symmetry ($U(1)$) known as a gauge symmetry). However, it was not known how to describe  the interactions of strongly interacting particles. Global symmetry were recognized as an important tool in their classification but there was the nagging feeling that something fundamental was missing. The bootstrap hypothesis of Chew\cite{chew} was proposing an entirely different approach to address the strong interaction of hadrons- which meant no role for perturbative field theory. This was disturbing since Quantum Field Theory (QFT) was so successful in QED. 

The story of weak interaction was even more chaotic. While V-A current current theory was so successful in describing some of the observations it was not a fullfledged QFT, where one could calculate higher order processes. This was even more unsettling since unlike strong interactions which had coupling constant of order unity, experiments had established that weak forces were characterized by  much smaller couplings (e.g. $G_F \sim 10^{-5}$ GeV$^{-2}$). Besides, it was unclear, where the dimensionality of GeV$^{-2}$ in $G_F$)  had its origin.

\subsection{Symmetries in weak interactions and the W boson}
There were several tell-tale signs of something deep and beautiful in the observations of weak interactions that can be summarized as:

\begin{itemize}

\item  The weak interactions of both the leptons  known in the sixties i.e. $e, \mu, \nu_e, \nu_\mu$ and hadrons without strangeness possessed could be parameterized in terms of a strength  $G_F$. This suggested a kind of unity among  different forms of weak interactions and perhaps  an underlying symmetry.

\item The V-A Lorentz character of the weak interaction is similar to the pure vector nature of  electromagnetic interactions embodied in QED. Suspicions were raised in the early sixties by Schwinger and Glashow~\cite{gl} that perhaps there was a local symmetry behind the weak forces. We  elaborate on this below.

\item The mass dimension of the current -current form of weak interactions suggested   that  just as the photon mediates electromagnetism, there might be a vector boson, called the $W$ boson, that mediates  weak forces. In fact weak mediator bosons in the GeV mass range were contemplated by Lee and Yang~\cite{ly} in several of their papers.

\end{itemize}

A dominant question in the late sixties was whether there is an ultimate theory of weak forces that embodies all of the above listed properties . As noted earlier, it was already realized that there is a similarity between the quark currents generating symmetries of strong interactions and the leptonic weak currents, as note earlier. Could these leptonic symmetries form a Lie algebra of a symmetry group that would ultimately unify all  weak processes into a single theory ? A  related question was, whether the symmetry could be a local symmetry as in QED ? There were several basic obstacles to this point of view : first  is that the photon is massless (which implies  the observed fact that electromagnetic forces are long range forces unlike the weak forces which appears to operate only at a point (e.g. inside a nucleus in nuclear beta decays). Second, weak processes seem to  involve change of electric charge of participating constituents forming thereby a  current unlike to the electromagnetic current, which is electrically neutral. On the other hand, considering  the Lie algebra of the weak currents, we also find a neutral current that must  participate in the weak interactions i.e.
\begin{eqnarray}
[ {\cal J}^+_0, {\cal J}^-_0]={\cal J}^0_0
\end{eqnarray}

There were two messages in this observations: a discouraging message  that at the time, there was no evidence for any form of neutral current interactions in nature in the 1960s and an encouraging part  that theories with non-abelian  local symmetries like in Eq. (3) were already constructed in mid 1950s by Yang and Mills, which extended the possibility of local symmetric theories to the domain of weak interactions. These theories also provided a way to explain the observed universality of weak interactions. Such theories would also lead to the existence of charged vector bosons (as well as a neutral spin one boson) coupling to weak currents as was being contemplated in the phenomenological W-boson theories of weak interactions. The ``bad news" however was that the Yang-Mills gauge invariance demanded that W-bosons be massless, just like the photon which is a major road block to extending  gauge theories to the domain of weak interactions.

\subsection{Constructing massive gauge theories}
The idea that non-abelian gauge theories could be used to describe weak interactions was considered so attractive that several attempts were made in the late sixties and early 70's to take the non-abelian gauge theories and ``cure" its mass problem by simply adding a mass term for the vector bosons as a symmetry breaking term, to see what kind of theories they led to. Such theories, unlike massive QED were found to be non-renormalizable i.e. had more infinities than could be absorbed by the process of standard renormalization procedure and therefore unsuitable for being a consistent and predictive  field theory of weak interactions. Unlike QED, such theories could not make any testable predictions.

Alternative techniques where vector boson masses in Yang-Mills theories were not put in by hand but were generated in a different way were  discovered in the mid sixties through the works of Brout, Englert, Higgs, Guralnik, Hagen and Kibble~\cite{behghk}. They used the property that when local symmetries are broken by vacuum rather than by explicit terms in the Lagrangian, that also leads to nonzero mass for the gauge vector bosons. This process is known as the spontaneously broken gauge theories. It was however not known in the sixties, whether these models were renormalizable or not. In the early seventies, it was shown in a ground breaking paper by Gerhard 't Hooft ~\cite{thooft} that use of  gauge freedom inherent in these theories makes it possible to explicitly demonstrate that the spontaneously broken gauge theories are indeed renormalizable and therefore expected to lead to theories where quantum corrections are finite and calculable. This major step made it meaningful to compare predictions from such theories with experimental results. The 't Hooft observation thereby revolutionized the field of weak interaction.

\section{Glashow-Weinberg-Salam (GWS) model of weak interaction}
Although it was not known in the sixties whether gauge theories of weak interactions would be renormalizable, that did not stop Glashow, Weinberg and Salam\cite{gws} from proposing a gauge theory based on the $SU(2)_L\times U(1)_Y$ group that at the lowest order had the right properties to describe the weak interactions involving known charged current and which is now referred to as the standard model of electroweak (EW) interactions. This model provides a unified description of both weak and electromagnetic interactions and has been thoroughly confirmed by experiments, the latest being the discovery of the Higgs boson at the LHC. Below, we give a brief overview of some of the symmetry aspects of the model. We omit the color quantum number\cite{owg} in our discussion since that does not pertain to weak interactions. Under the weak $SU(2)_L\times U(1)_Y$ group, the fermions of one generation are assigned as follows:(the numbers within the parenthesis denote the weak isospin $I$ and hypercharge $Y$ quantum numbers)
\begin{eqnarray}
Quark ~doublets:~ Q_L~= \left(\begin{array}{c} u_L\\d_L \end{array}\right) (1/2, 1/3);\\ \nonumber
Lepton ~doublets:~ L~= \left(\begin{array}{c} \nu_L\\e_L \end{array}\right)\equiv (1/2, -1);\\ \nonumber 
Right handed ~singlets~: u_R(1,4/3); d_R(1,-2/3); e_R(1,-2)
\end{eqnarray}
where $u, d, \nu, e$ are the up, down quarks and the neutrino and electron fields respectively. The subscripts $L, R$ stand for the left and right handed chiralities of the corresponding fermion fields. 
The electric charge of the particles is given by $Q=I_3+\frac{Y}{2}$.
There are four gauge bosons $W^{\pm}_\mu, W^3_\mu, B_\mu$ associated with the four generators of the gauge group, The interactions of these gauge fields with matter (the quarks and leptons) are determined by the symmetry of the theory and lead to the current-current form for weak interactions via exchange of the $W^\pm$ gauge boson. Before symmetry breaking, the gauge bosons and fermions are all massless. The masslessness of the gauge bosons is analogous to that of the photon  in QED. Since  fermion mass terms which are bilinears of the form $\bar{\psi}_L\psi_R$  connect the left and right chirality states of the fermion, such terms are also forbidden by gauge invariance since the left chiral states of fermions in SM are $SU(2)_L$ doublets whereas the right chirality ones are singlets. 

To give them mass, we adopt the model of spontaneous breaking of gauge symmetry by including in the theory scalar Higgs fields, $\phi(1/2, 1)$ which transform as doublets of the gauge group (or weak isospin 1/2). This allows Yukawa couplings of the form $\bar{Q}\phi d_R$, $\bar{\psi}_L\phi e_R$ and $\bar{Q}\tilde{\phi} u_R$ where $\tilde{\phi}=i\tau_2\phi^*$ ($\tau_{1,2,3}$ denote the three Pauli matrices). If the gauge symmetry is broken by giving a ground state value to the field $\phi$ as $<\phi >=\left(\begin{array}{c}0\\ v\end{array}\right)$, this gives mass not only to all the fermions in the theory but also to the gauge bosons $W^\pm$ and to $Z\equiv \cos\theta_W W_3+\sin\theta_W B$ where $\theta_W=tan^{-1}\frac{g^\prime}{g}$ is the Weinberg angle, which is related to the gauge couplings of the weak and electromagnetic forces. Note $<\phi>$ leaves one gauge degree of freedom unbroken i.e. $Q=I_3+\frac{Y}{2}$ (where $I_3$ denotes the third weak isospin generator of the $SU(2)_L$ gauge group)  since $Q<\phi>=0$. $Q$ can be identified as the electric charge and given the quantum numbers assignments to different particles, reproduces the observed electric charges of the particles of SM.  

\subsection{Key predictions of the standard model and its experimental confirmation} The standard model has many predictions for the gauge boson sector, as well as for the gauge interaction to fermions. To see this,  we note that  aside from the fermion sector, the model has three parameters that define the masses and interactions of the gauge bosons i.e. the two gauge couplings $g, g^\prime$ and the Higgs vacuum expectation value(vev)  $v$ in the ground state of the theory .  Since electromagnetic interactions are generated out of the gauge interactions of the model, electric charge can be expressible in terms of the gauge couplings $g, g^\prime$ and the formula is:
\begin{eqnarray}
\frac{1}{e^2}=\frac{1}{g^2}+\frac{1}{g^{\prime 2}}
\end{eqnarray}
In terms of the Weinberg angle $\theta_W$, we have $g=e~ $cosec~$ \theta_W$ and $g^\prime = e $~sec~$ \theta_W$. The Fermi coupling is expressed in terms of the same parameters as : $\frac{G_F}{\sqrt{2}}= \frac{1}{4v^2}$. This leaves us with  $\sin^2\theta_W$ as the only unknown parameter. This parameter determines the structure of the neutral current couplings predicted by this theory. We also have the masses of the $W^\pm$ and the $Z$ boson that can be expressed as:
\begin{eqnarray}
M_W=\frac{ev}{\sqrt{2}}; M_Z=\frac{ev}{\sqrt{2}\cos\theta_W}
\end{eqnarray}
The form of the weak interaction Hamiltonian following symmetry breaking is :
\begin{eqnarray}
{\cal H}_{I}~=~\frac{g}{2\sqrt{2}}W^+_\mu[\bar{P}\gamma^\mu(1-\gamma_5)V_{CKM}N+\bar{\nu}\gamma^\mu(1-\gamma_5)\ell~+~h.c.]\\ \nonumber
               + \frac{e}{\sin\theta_W cos\theta_W}Z_\mu \sum_i \bar{\psi}_i\gamma^\mu(I_{3}-Q\sin^2\theta_W)\psi_i\\\nonumber
                +h\sum_i \frac{m_i}{M_W}\bar{\psi}_i\psi_i
                \end{eqnarray}
                where $\bar{P}=P^\dagger\gamma_0$; $P$ is a column vector given by $P^T=(u,c,t)$, $N=(d, s, b)$ , $\ell=(e,\mu,\tau)$ and $\psi_i$ summed over all the fermions of SM; $h$ is the Higgs field; $V_{CKM}$ is the quark mixing matrix. Armed with this information, one can now test the model in experiments.
                
The first confirmation of the GWS model came from the discovery of the neutral weak interactions between neutrinos and quarks and electrons. Many observables were expressed in terms of only one unknown $\sin^2\theta_W$ and remarkably, all neutral current observations including quantum corrections\cite{marciano}  were fitted with only this one parameter, whose value was determined to be $\sin^2\theta_W=0.233$. Given the value of $\sin\theta_W$, the model predicted the value of the masses of the $W$ and $Z$ bosons to be $80$ GeV and $91$ GeV and soon experiments \cite{WZ} discovered these particles with precisely the masses predicted by the standard model. Furthermore, the model predicts that flavor changing neutral current effects such as $K_L\to \mu^+\mu^-$, $K_L-K_S$ mass difference are suppressed to the observed level naturally without extra inputs or fine tuning of any parameters.

The standard model has a very interesting feature as far as including the violation of matter-anti-matter symmetry (or CP).  CP violation in the kaon decays were observed in 1964\cite{fitch} in the decays of the $K$ meson the same particle whose decay property was also at the root of the discovery of parity violation.  If the standard model has to provide a realistic description of observed phenomena, it must incorporate CP violation. Note that incorporating CP violation requires the presence of a complex phase in some interaction of the fermions in a field theory. Gauge symmetry ensures that neutral current interactions of the model conserve CP. For the Higgs doublet similarly, the couplings are automatically CP conserving. The only place for a complex phase is therefore the $W^\pm$ interaction with the left handed (or $V-A$) current. In particular the way it could appear is in the mixing between the different quarks in the weak current. For instance, if the weak current is written as $\bar{u}\gamma^\mu (1-\gamma_5)(d \cos\theta+e^{i\delta} s \sin\theta)$, the complex phase $\delta$ will create asymmetry between the particle and anti-particle weak decays. However, since complex fermion fields can be redefined to absorb a phase from the interaction, introducing a physical CP phase that cannot be removed from the theory by redefinition of fields is highly nontrivial and puts constraints on the model. In fact very early in the development of gauge theories, it was shown that with two generations of quarks, standard model cannot incorporate CP violation \cite{rnm}. It was subsequently shown in a prescient paper by Kobayashi and Maskawa\cite{km} that if there are three generations of quarks, one can have a single nontrivial phase in the theory and theory will violate CP. This proposal was made before there was any evidence for the third generation of quarks, which were subsequently discovered in experiments at Fermilab\cite{bquark} as well as the third generation charged lepton, the $\tau$ lepton, discovered at SLAC\cite{perl}. This Kobayashi-Maskawa theory was subsequently confirmed by many experiments\cite{babar} and replaced the old superweak theory of CP violation\cite{wolf}.

Final confirmation of the Glashow-Weinberg-Salam theory has come recently from the discovery of a Higgs like particle with a mass of 125 GeV at the Large Hadron Collider. Unlike the masses of the $W$ and $Z$ bosons, the mass of the Higgs boson, $m_h$ is an arbitrary parameter of the standard model and  not predicted by the theory. However, comparisions of the measured electroweak parameters with calculations from SM of the same parameters indicates  that $m_h$ must be in the 115 to 150 GeV range.There is a  therefore, a strong belief that the newly observed 125 GeV particle is the Higgs boson. The Higgs boson is  highly unstable which makes it difficult to detect among the trillions of end products such as quark jets, leptons, photons that are produced in such an energetic collision of proton beams. By the same token, since the standard model predicts very definite rates for its various decay modes, once the Higgs boson is detected, measurement of its various decay rates to different final states can be used to nail down down its true identity and finally confirm that the 125 GeV boson is indeed the Higgs boson of the standard model. Another reason for the euphoric excitement among particle physicists that followed the LHC results is that once this particle is confirmed, this will be the first, possibly not the last  elementary scalar particle to be discovered in nature. In fact,  there have been conjectures that similar scalar particles may also be at the heart of the cosmic expansion at the beginning of Big Bang known as inflation. Therefore, studying properties of the Higgs boson  and search for other Higgs-like bosons are going to be the focus of  future experimental and theoretical activity in the field for many years to come.

\subsection{Puzzles of the standard model}

The standard model is extremely successful in describing the  observations both in the domains of weak as well as electromagnetic interactions at low energies. When supplemented by the gauge theory of color, the $SU(3)_c$, it accounts for many of the observed properties of strong interactions, e.g  its low energy symmetries, its behavior at high energies etc.
This theory involving color  is known as Quantum Chromodynamics and is accepted as the theory of strong interactions. The standard model still however faces a number of conceptual as well phenomenological problems.

At the phenomenological level,  several observations cannot be accommodated within the standard model:

\begin{itemize}

\item Neutrinos are massless in the standard model; however, observations of the oscillation of neutrinos coming from the Sun and the cosmic rays has established beyond any doubt that neutrinos have mass. Furthermore during the past 15 years since the discovery and confirmation of neutrino oscillations, measurements of various neutrino mass differences and the mixings between various generations of neutrinos have been measured. They are  similar to the mixings among different  quarks of same electric charge which have been measured in many different accelerator as well as low energy experiments. The mixings among neutrinos were already discussed in the early sixties by Pontecorvo and Maki, Nakagawa and Sakata\cite{pmns}. The neutrino oscillation measurements have raised  new puzzles since the pattern of mixings among the neutrinos is very different from those among quarks~\cite{bilenky1}.

\item  The existence of dark matter in the universe has been confirmed through a variety of cosmological and astrophysical observations. The fact that it is ``dark" implies that there must be a new kind of particle that has no electromagnetic as well as strong interactions; on the other hand the fact occurs with a certain abundance requires that it has to have some other kind of interactions that allows this to happen. There seem to be no particles in the standard model that can play the role of dark matter and one must look beyond.

\item Finally, fundamental theories of matter predict that there must also be anti-matter accompanying matter. The early universe is therefore likely to contain matter and anti-matter in equal abundances. However, all observations on large as well as small scales in the universe  have found no evidence for anti-matter - only matter and that the density of matter is only a billionth of the photon density in the current epoch of the universe. There is no mechanism to understand this tiny asymmetry between matter and antimatter within the framework of the standard model which implies that there must be new physics beyond.

\end{itemize}

There are also conceptual difficulties with the standard model, in that it does not provide a prescription for all observations in terms of a few basic inputs  without arbitrary adjustments of the input parameters, which could, in principle be a requirement for a truly fundamental theory. 


\noindent (i)  Foremost among the issues is the so called gauge hierarchy problem, which states that if there is no new physics just beyond the standard model except for the onset of quantum gravity, whose scale is the value of the Planck mass of $M_P\simeq 10^{19}$ GeV,   quantum effects are then most likely to set the mass of the Higgs boson to the very same scale of $10^{19}$ GeV, rather than 125 GeV where it  seems to be and which would be expected if it is to describe the scale of weak interaction. The nature of any new physics beyond the standard model must not only keep the Higgs mass from shooting up sky high, but it must also keep it near 125 GeV after quantum effects of that additional contributions are included. At present an attractive theory that seems to go a long way to solving this problem is ``supersymmetry"(or SUSY)\cite{haber}. A widely held belief is that the Large Hadron Collider should also provide evidence for supersymmetry. However so far it has failed to do so. In any case, the narrative of weak interaction we are focussing on in this review is not likely to be much affected by the presence and therefore, we will not discuss the fascinating subject of supersymmetry and its deep connection to gravitational forces.

\noindent (ii) Among other puzzles of the standard model, we can list  the so called "flavor problem" which proposes to confront the pattern of fermion masses and mixings in terms of  fundamental symmetries;  the strong CP problem which arises from a non-perturbative property of the  QCD theory of strong forces, which introduces an arbitrary strength strangeness conserving time reversal violation into the theory for which there is no evidence. Clearly understanding this needs an extension of the standard model.

\noindent (iii) Finally,  why are the weak interactions left handed rather than mirror symmetric as their other cousins, the strong, electromagnetic and gravitational forces. Could it be that at some fundamental level, weak interactions do really conserve mirror symmetry (parity) and fail to do so at low energies where they  break parity symmetry. Two classes of suggestions have been proposed in response to this open issue. 

(a) One class of models was first suggested in the famous Lee-Yang paper that proposed parity violation in weak interactions, according to which our universe must be accompanied by a parallel universe that contains identical matter and forces and under parity operation, particles in our universe transform into particles of the other universe where weak interactions are right handed. So when we include both universe together in our description of forces and matter, there is no mirror asymmetry. Since the particles of the other universe have forces to which particles of our universe are not sensitive, it looks like weak interactions break mirror symmetry. These classes of models are difficult to probe  experimentally since the so called ``mirror particles'' that are part of the mirror reflected sector of our universe have no interactions with the force carriers of our universe (except for gravity) i.e. photons, gluons, W and Z bosons, etc.

(3b) The other class of suggestion arose after the discovery of gauge theories and proposes\cite{mp} that unlike the standard model that gauges asymmetrically the left handed matter, the ultimate theory has gauge forces that couple to both left as well as right handed matter (i.e. quarks and leptons) and we only see the interactions of left-handed matter because the analog of the $W$ boson (the $W_R$ boson) is much heavier than the known $W$ or $W_L$ boson thereby suppressing its four-Fermi interaction at low energies. We will see below that one may interpret the observation of neutrino mass as the first indication that such right handed forces exist in Nature. This leads to many new effects in the domain of weak interactions which are currently being actively searched for in experiments including the LHCand we discuss it in a separate section below.


\section{Neutrino mass and going beyond the standard model}
The first evidence for  physics beyond the standard model came with the discovery of neutrino mass in 1998. 

\subsection{Discovery of the neutrino mass} 
For a long time, it was thought by many particle physicists that neutrinos are massless and travel with the speed of light. However, there were theoretical suspicion among a few e.g. the practitioners of the so-called left-right symmetric theories that I describe below that it is more natural for neutrinos to have mass. From time to time, there were also experimental hints for neutrino mass. However, first serious evidences started appearing from experiments on neutrino oscillation. In particular,  from an experiment performed for many years by Ray Davis in a gold mine in Homestake, South Dakota where he was trying  to detect  neutrinos emitted from the sun. The energy in sunshine comes from a nuclear weak process in the core of the Sun where four protons and two electrons combine to eventually produce a Helium nucleus together with two electron neutrinos with energies in the KeV to about 14 MeV range. The expected solar flux of neutrinos was  predicted by John Bahcall~\cite{bahcall}. Davis's initial results in mid sixties indicated that he observed about a third of the expected number of neutrinos\cite{davis}. Davis devoted most of his life to this experiment and always continued to get similar results. One way to understand his results is to assume that neutrinos have mass and oscillate from one species i.e. $\nu_e$'s emitted from the core of the Sun to another say $\nu_\mu$ in its journey  to the Earth. Since the Davis experiment could not detect muon neutrinos, that would lead to a reduction of the solar flux. It is important to emphasize that for such oscillation to occur, neutrinos must have finite mass in contrast to the prediction of the standard model.

These results were confirmed in a brilliant experiment conducted in the Kamioka mine in Japan in 1998\cite{sk} where they not only observed a reduction in the solar neutrino flux (as Ray Davis did) but also in the flux of cosmic ray neutrinos (which consist dominantly of muon neutrinos) from the atmosphere. This established that not only electron neutrinos oscillated but also so did muon neutrinos. This was soon followed by searches for solar neutrino oscillations in an experiment in Sudbery mine in Canada (known as the SNO experiment\cite{sno}) and that for the reactor anti-neutrinos in Japan, known as KamLand experiment\cite{kamland}. All of them have given a very clear picture that neutrinos have mass and they mix much like the quarks do , although with very different mixing angles.

Coming to grips with solar neutrino oscillations uncovered a new phenomenon involving neutrinos, which is the analog of refraction of light  passing through glass. This arises from the scattering of neutrinos off matter as it passes through them. This is known as Mikheyev-Smirnov-Wolfenstein effect~\cite{msw} and is a phenomenon always present  to a small or large extent when neutrinos pass through matter.

Direct searches for neutrino masses by studying its effect on the endpoint of the energy spectrum of electrons emitted in beta decays  of nuclei such as tritium have also continued, the latest being the KATRIN~\cite{katrin} experiment in Germany, however no positive signal has yet been observed .

\subsection{Implications for theory}

The fact that they have spin half  means that massless neutrinos produced in one helicity state remain frozen in that state for ever. It was this observation that formed the foundation of the V-A theory of weak interactions. The argument is as the follows. A massless neutrino obeys invariance under the transformation $\psi \to \gamma_5\psi$ for the neutrino field. If the same transformation is assumed to hold for all fermion fields participating in weak interactions, it automatically implies that only left handed components of the fields can have weak interactions and hence the Lorentz structure of weak 
interactions must be V-A type and must maximally violate parity invariance. This of course raised the parity puzzle of why weak interaction is the only force that violates parity.

The intimate connection between the parity puzzle of weak interaction and neutrino mass becomes clear in the left-right symmetric models of weak interactions\cite{mp}. Since left-right symmetric models of weak interactions necessarily have right handed neutrinos, neutrinos automatically acquire masses through the same Higgs mechanism that gives mass to the quarks and charged leptons. The conceptual problem with this notion is that one would naively expect the neutrino masses to be of the same order as any of the quark or lepton masses.
This is contradicted by experiments that discovered  neutrino oscillations,  which have established that neutrino masses must be less than an electron volt.  This is many orders of magnitude smaller than expectations based on this naive reasoning. 

\subsection{Left-right symmetric models of weak interactions- a solution to the parity and neutrino mass problem}
In this subsection, we discuss how one can provide simultaneous resolution of the parity puzzle as well as neutrino mass  in the left-right symmetric (LRS) extension~\cite{mp} of the standard model. The gauge group of the LRS model is: $SU(2)_L\times SU(2)_R\times U(1)_{B-L}$ including discrete parity symmetry and with fermion assignments given by:
\begin{eqnarray}
Q_{L,R}~= \left(\begin{array}{c} u\\d \end{array}\right)_{L,R} (1/2,0, 1/3)~or (0,1/2,1/3);\\ \nonumber \psi_{L,R}~= \left(\begin{array}{c} \nu\\e \end{array}\right)_{L,R}\equiv (1/2,0, -1);~or ((0,1/2,-1)
\end{eqnarray}
where the numbers in the parenthesis refer to the $SU(2)_L$, $SU(2)_R$ and $U(1)_{B-L}$ quantum numbers.
Because under parity inversion, left-handed fermions go to right handed fermions, the above assignment is parity symmetric. The resulting weak interaction Lagrangian is given by:
\begin{eqnarray}
{\cal L}_{wk}~=~i\frac{g}{2}\left(\bar{Q}_L\vec{\tau}.\vec{W}^\mu_L\gamma_\mu Q_L+\bar{\psi}_L\vec{\tau}.\vec{W}^\mu_L\gamma_\mu \psi_L\right)~+~L\leftrightarrow R
\end{eqnarray}
 If under parity inversion,  $W_L$ transforms to $W_R$, this Lagrangian conserves parity. However once symmetry breaking is turned on, the $W_R$ will acquire a higher mass and introduce parity violation into low energy weak interaction. The effective weak interaction Hamiltonian below the $W$ boson mass can be written as:
\begin{eqnarray}
{\cal H}_I~=~\frac{g^2}{2M^2_{W_L}}({\cal J}^{+,\mu}_L{\cal J}^{-}_{\mu,L})+\frac{g^2}{2M^2_{W_R}}({\cal J}^{+,\mu}_R{\cal J}^{-}_{\mu,R})~+~h.c.
\end{eqnarray}
When $m_{W_R}\gg m_{W_L}$, this weak interactions violate parity conservation almost maximally as the right handed current effects are suppressed by a factor $\frac{m^2_{W_L}}{m^2_{W_R}}$. Given the accuracy in the tests of V-A theory at low energies therefore, one can obtain limits on the mass of $W_R$ boson~\cite{beg}. There are more recent detailed analysis of these limits which put a lower bound on the mass of $W_R$ around 2.5 TeV. In addition, the theory can also accommodate left-right mixing terms, which also have limits on them.

The connection between spontaneous parity violation and small neutrino masses was first noted in \cite{ms}. It was pointed out that if the Higgs field that breaks the $SU(2)_R\times U(1)_{B-L}$ symmetry down to $U(1)_Y$ of the standard model is so chosen that while giving mass to the $W_R$ and the $Z^\prime$ boson, it also gives a Majorana mass $M_R$  to the right handed neutrino, then the mixed left and right handed neutrino mass matrix acquires the form
\begin{eqnarray}  
{\cal M}_{\nu, N}~=~\left(\begin{array}{cc} 0 & m_D\\ m^T_D & M_R\end{array}\right)
\end{eqnarray}
where each of the entries are $3\times 3$ matrices. Diagonalisation of this mass matrix leads to a mass formula for the light neutrinos of the form
\begin{eqnarray}
{\cal M}_\nu~=~-m_D M^{-1}_R m^T_D
\end{eqnarray}
Note that since $M_R$ corresponds to the right handed neutrino scale, it is not restricted by the standard model physics and can be large whereas $m_D$ is proportional to the scale of standard electroweak symmetry breaking and therefore of the same order as the quark-lepton masses. Thus by making $M_R$ large, we can obtain a very tiny neutrino mass matrix. This is known as the seesaw mechanism~\cite{seesaw}. In the context of the left-right symmetric model, since $W_R$ as well as $N_R$ mass are proportional to the same scale, the smallness of the neutrino mass and suppression of the V+A current interactions are proportional to each other. In the limit that neutrino mass goes to zero, the weak interactions become pure V-A as in theories with $\gamma_5$ invariance. Thus the left-right seesaw models provide a connection between the path from massive to massless neutrino and the path from some V+A interactions to pure V-A weak interaction.

An important implication of the seesaw mechanism is that neutrinos are Majorana fermions i.e. they are their own anti-particles. As noted before since neutrinos are supposed to carry lepton number $L=1$, neutrino and the anti-neutrino are supposed to have opposite lepton number; however, if neutrinos is its own anti-particle, this will clearly break lepton number by two units. One should search for experimental verification of this possibility, which will provide a first key test of the seesaw mechanism.

\subsection{Majorana neutrinos and neutrino-less double beta decay}
A subject s under a great deal of scrutiny  now-a-days is the resolution of whether neutrinos are  Majorana or Dirac particles. As noted above, one way to understand the tiny masses for neutrinos is via the seesaw mechanism which leads to Majorana neutrinos, a  the neutrino mass term that violates lepton number symmetry by two units. One way to test this possibility is to search for nuclear as well as rare processes where such violation of lepton number occurs. It was suggested by Furry in 1939~\cite{furry} that there are certain nuclear rare decay processes where this could be tested. He pointed out that there are beta unstable nuclei (A, Z) in nature which are kinematically forbidden from decaying into
(A, Z+1) via neutrons decaying to $p+e^-+\bar{\nu}_e$ rather they can decay with two neutrons decaying to two protons whereby we have $(A,Z)\to (A, Z+2)+e^-e^-\bar{\nu}_e \bar{\nu}_e$ via second order weak interaction. Furry pointed out that if neutrinos are their own anti-particles, the two neutrinos emitted in the above second order weak interaction process ``annihilate" each other leading to the so-called neutrino-less double beta decay process where $(A,Z)\to (A, Z+2)+e^-+e^-$ and no neutrinos. The mass difference between the initial and final nuclei is then equally divided between the two electrons giving mono-energetic neutrinos with $E_e= \frac{1}{2} (m_{(A,Z)}-m_{(A, Z+2)})$ for each electron i.e. a spike in the plot of number of events vs total electron energy at the end point of the conventional two neutrino double beta decay. The two neutrino double beta decay was observed by Elliott , Hahn and Moe in 1995~\cite{moe} in the decay of Selenium nucleus ($^{82}$Se$\to$ $^{82}$Kr). The neutrino-less double beta decay has been searched for many years and is being searched for currently with impressive progress  in $^{76}$Ge$\to$ $^{76}$Se, $^{136}$Xe$\to$ $^{136}$ Ba, $^{100}$Mo $\to$ $^{100}$Ru, etc. 

The observation of neutrino-less double beta decay ($\beta\beta_{0\nu}$) will be a discovery of fundamental significance.  Lepton and baryon number are two of the fundamental symmetries of the standard electroweak model. Searches for baryon number number violation have not been successful so far. Observation of neutrino-less double beta decay will establish that lepton number is violated by two units and will provide the first evidence one of the sacred conservation laws of the standard electroweak model is broken. This will be of major significance in our exploration of physics beyond the standard model\cite{bilenky}.

Furthermore,  since the neutrino contribution to neutrinoless double beta amplitude is proportional to its mass, a signal in $\beta\beta_{0\nu}$ could be a direct measure of the neutrino mass. There could however be other heavy particle contribution to this process e.g. contributions from right handed $W$-bosons in combination with right handed neutrino Majorana mass or lepton number violation in supersymmetric theories etc. The latter effects can however be tested using the Large Hadron Colliders, thus making the search for 
$\beta\beta_{0\nu}$ of much broader interest in particle physics than just the neutrino mass area.

\subsection{Unsolved issues in neutrino mass physics} Studies of neutrino mass  have become a gold mine for searches of new physics beyond the standard model. Despite the fact that non-zero neutrino mass itself is evidence of new physics, there are observations in neutrino oscillation physics that call for new ideas. One glaring example of of a point of interest  is the diverse  pattern of mixing between quarks and leptons. As noted earlier, the different generations of quarks mix among each other when they participate in the weak interactions. This mixing is known as the Cabibbo-Kobayashi-Maskawa (CKM) mixing~\cite{km}. This mixing has the property that it is hierarchical in pattern i.e. the weak interaction Lagrangian is characterized by a weak mixing matrix which is a unitary matrix whose diagonal elements are its dominant component and off diagonal ones are small. On the other hand, the corresponding mixing matrix in the lepton sector is far from hierarchical. In fact, if we denote $\theta_{12}, \theta_{23}, \theta_{13}$ as the mixing angles between the different generations of leptons (denoted in the subscript), we have $\theta_{23}\sim 40^0$ in the lepton sector whereas it is about $3^0$ in the quark sector. Similarly, the mixing angle $\theta_{13}$ in the lepton sector is $9^0$ whereas that in the quark sector is less than a degree. These diversity suggests that new physics in the quark and lepton sectors could be very different. Yet many models have been proposed which unify quarks and leptons and reproduce the diverse patterns purely out of the dynamics of the model.
The full story behind these diverse patterns of mixing is however yet to be written.

Other unsolved issues pertain to the ordering of  masses of neutrinos. Defining the pattern of quark and charged fermion masses as normal (which means that higher generations have higher masses), current results from oscillations are fully consistent with the neutrino mass patterns being either normal or inverted. Experimental efforts are under way to settle this issue. Similarly, it is not known if there is CP violation in the lepton sector. If intuition based on the quark observations can be trusted, one should expect that there will be CP violation among leptons. There are conjectures that this leptonic CP violation in a seesaw framework could explain the matter -anti-matter asymmetry in the Universe, one of the fundamental mysteries of the standard model of particle physics.

Other new possibilities include the  existence of new species of neutrinos which do not participate in conventional weak interactions, the so-called sterile neutrinos. At present there are anomalies that provide hints for this possibility although it is far from being confirmed. The existence of sterile neutrinos if confirmed  will be another
major revolution in this field and might force many of the current laws to receive substantial revision.

\subsection{Applications to astrophysics and cosmology} With the progress of weak interaction theory, our understanding of the universe has also improved. This is a large topic and I will briefly mention a few examples. For instance, one of the observations about the universe today is its Helium content by mass, $Y_p\equiv \frac{f_{He}}{f_H+f_{He}}\approx 0.24$. In order to understand this observation within the Big Bang model of the universe, we first note that the expansion rate of the universe is given by the number of species both relativistic and non-relativistic which contribute to its energy content, which by Einstein's equation for matter coupled to gravity determine its expansion rate. To understand the formation of Helium, the first thing to note is that we need to have a steady content of neutrons and protons so that nuclear reactions can combine neutrons and protons to form Helium.  Second requirement is that  the temperature of the universe at the Helium formation epoch must be low enough so that the nuclear reaction chain through which Helium is formed, must be allowed. Take for example Deuterium which is a step through which proton and neutrons combine before they reach the stage of Helium formation. Thus Deuterium must form before, one can have Helium formation. Deuterium has a binding energy of nearly 2.2 MeV . So if the temperature if the universe is above 2.2 MeV, any deuterium formed will break and Helium formation will be hampered. To see how the first condition is satisfied, note that if weak interaction rates are fast, the number of neutrons and protons keep changing due to the reaction $\nu + n\to e^-+p$ and its inverse. In order for Helium formation to start, weak interactions must be slow enough so that these reactions fall out of equilibrium (i.e. cannot keep pace with expansion rate and will stop) keeping the neutron proton ratio fixed.  It turns out that the weak interaction coupling is such that weak interactions go out of equilibrium around one MeV. This temperature being  less than the Deuterium binding energy, deuterium can form without disintegrating, thereby facilitating the formation of Helium. This is a very rough picture but captures the essence of the idea.  Taking Boltzmann distribution for the neutrons and protons at the freeze-out when weak process $\nu + n\to e^-+p$ stops , one can predict a value for the Helium abundance observed. These ideas which form the basis of the modern understanding of the Helium, Deuterium and Lithium abundances today, have been refined and used to derive the number of neutrino species in particle physics and our understanding of weak interactions play a crucial role in this\cite{gary}.

There are other areas of astrophysics where weak interactions are also believed to play an important role. For instance, one of the puzzles of the Big Bang theory is to understand the abundance of  heavy elements on Earth. A plausible source is supposed to be the supernova explosions where in outer sphere during the explosion, neutrino interactions generate a neutron rich environment causing the formation of heavy elements via successive complex nuclear reactions. Here again, it the weak interaction forces which play a key role. Other possibilities include the reason for the supernova explosion itself which could have possibly been caused by the coherent neutrino neutral current scattering against heavy nuclei. Thus further understanding of weak interaction phenomena can shed light on these and other cosmological and astrophysical issues.

\section{Summary} The field of weak interactions has been a vibrant field for the past century and with the advent of its description in terms of the standard model, has represented a major triumph of perturbative quantum field theory. Through weak interactions, we have been able to get a glimpse of the underlying symmetries of quarks and leptons and their realization in nature. Local symmetries have now become the cornerstones of these theories. Weak interactions may have taught us about the origin of all mass in the universe  i.e. the so-called Higgs mechanism. The second phase in the history of weak interactions likely began with the discovery of neutrino oscillations more than decade ago. If the seesaw model for neutrino masses is correct, it will represent another source of mass  in the universe. This, in turn,  is opening up the possibility of new symmetries of nature, whose confirmation will have profound implications in our journey to explore the mysteries of the universe. This process may also reveal the secrets behind some of the deep unsolved puzzles such as the origin of matter in the universe including the dark component of the matter which exceeds the visible component by a factor of five. The quest for an ultimate theory of nature continues, and weak interactions may be leading the way.

\section*{Acknowledgement} This work has been supported by the National Science Foundation grant No. PHY-0968854. I  thank E. Henley and Tom Ferbel for  reading the manuscript and suggesting many improvements.

\end{document}